\documentclass[conference]{IEEEtran}
\IEEEoverridecommandlockouts
\usepackage{cite}
\usepackage{amsmath,amssymb,amsfonts}
\usepackage{algorithmic}
\usepackage{graphicx}
\usepackage{textcomp}
\usepackage{xcolor}
\usepackage{url}

\def\BibTeX{{\rm B\kern-.05em{\sc i\kern-.025em b}\kern-.08em
    T\kern-.1667em\lower.7ex\hbox{E}\kern-.125emX}}

\DeclareRobustCommand{\erase}{\bgroup\markoverwith{\textcolor{red}{\rule[.5ex]{2pt}{0.4pt}}}\ULon}

\begin{document}

\title{Quantum Local Search for Traveling Salesman Problem with Path-Slicing Strategy
\thanks{
    This work was performed for Council for Science, Technology and Innovation (CSTI), Cross-ministerial Strategic Innovation Promotion Program (SIP), ``Promoting the application of advanced quantum technology platforms to social issues'' (Funding agency: QST).}
  
}
\author{
\IEEEauthorblockN{
    Chen-Yu Liu \IEEEauthorrefmark{1}\IEEEauthorrefmark{2}\IEEEauthorrefmark{3}, 
    Hiromichi Matsuyama\IEEEauthorrefmark{1}\IEEEauthorrefmark{4},
    Wei-hao Huang\IEEEauthorrefmark{1}\IEEEauthorrefmark{5},
    Yu Yamashiro\IEEEauthorrefmark{1}\IEEEauthorrefmark{6},
}

\IEEEauthorblockA{\IEEEauthorrefmark{1} Jij Inc., 1-4-6 Nezu, Bunkyo, Tokyo 113-0031, Japan}
\IEEEauthorblockA{\IEEEauthorrefmark{2}Graduate Institute of Applied Physics, National Taiwan University, Taipei, Taiwan}

\IEEEauthorblockA{Email:\IEEEauthorrefmark{3} d10245003@g.ntu.edu.tw, \IEEEauthorrefmark{4} h.matsuyama@j-ij.com, \IEEEauthorrefmark{5} w.huang@j-ij.com, \IEEEauthorrefmark{6} y.yamashiro@j-ij.com
}
}

\maketitle

\begin{abstract}
We present novel path-slicing strategies integrated with quantum local search to optimize solutions for the Traveling Salesman Problem (TSP), addressing the limitations of current Noisy Intermediate-Scale Quantum (NISQ) technologies. Our hybrid quantum-classical approach leverages classical path initialization and quantum optimization to effectively manage the computational challenges posed by the TSP. We explore various path slicing methods, including k-means and anti-k-means clustering, to divide the TSP into manageable subproblems. These are then solved using quantum or classical solvers. Our analysis, performed on multiple TSP instances from the TSPlib, demonstrates the ability of our strategies to achieve near-optimal solutions efficiently, highlighting significant improvements in solving efficiency and resource utilization. This approach paves the way for future applications in larger combinatorial optimization scenarios, advancing the field of quantum optimization.
\end{abstract}


\section{Introduction}





Quantum computing shows promise as an innovative technology, providing potential computational benefits for various applications, particularly in combinatorial optimization \cite{isingformulation, abbas2023quantumopt}. Among these applications, the Traveling Salesman Problem (TSP) stands out due to the intuitive clarity of its problems and the difficulty of solving them.  The TSP, which involves finding the shortest possible tour that visits a set of locations and returns to the origin, is notorious for its combinatorial explosion as the number of locations increases. Quantum computing is expected to tackle such problems more efficiently than classical computing by exploiting quantum phenomena such as superposition and entanglement.


However, the current era of Noisy Intermediate-Scale Quantum (NISQ) technologies presents challenges, primarily due to the limited number of qubits available on quantum devices. This limitation is a significant barrier as the TSP and similar combinatorial optimization problems typically require a large number of variables to be encoded for quantum processing. The scaling of problem size with the number of qubits directly impacts the feasibility and efficiency of quantum algorithms.

Previous studies have developed qubit-efficient methods to address these limitations, particularly for the TSP. 
These include using Tabu search for partitioning problems and quantum annealers to solve subproblems \cite{qls6}, the quantum phase estimation method with gate-based quantum computers \cite{tsp1}, and quantum local search methods such as D-Wave's Qbsolv \cite{tsp2}. Extending these techniques to general combinatorial optimization problems, the focus is on reducing the number of qubits required to encode a problem \cite{qls1, qls2, qls3, pqls, qls4, qls5}, thereby making the best use of the limited qubit resources available on NISQ devices. Quantum local search, in particular, adapts the local search strategy—commonly used in classical optimization—to the quantum context, allowing for iterative improvement of solutions by exploring a quantum state space.



Despite the advances in quantum local search, there are intrinsic limitations when applied to complex problems like the TSP. Traditional quantum local search approaches might not efficiently handle the unique challenges posed by TSP, such as the need for encoding cyclic paths and managing large combinatorial spaces without escalating the qubit requirements exponentially. Therefore, a specific adaptation of quantum local search tailored to the TSP's requirements is essential for improving solution quality and computational efficiency.


In this work, we introduce novel path-slicing strategies \cite{ps1} integrated with quantum local search to tackle the TSP. In Sec. \ref{sec:method}, we detail our methodology, combining classical path initialization with quantum optimization. In Sec. \ref{sec:result}, we present our results, demonstrating scalability and efficiency, and discuss potential future applications. Finally, in Sec. \ref{sec:conclusion}, we summarize our findings and their significance.


\section{Path-slicing as a quantum local search method}
\label{sec:method}

In this section, we will introduce our proposed path-slicing method for quantum local search. Followed by a description of path initialization method, and the different strategies for path slicing. 

\begin{figure}[ht]
\centering
\includegraphics[scale=0.25]{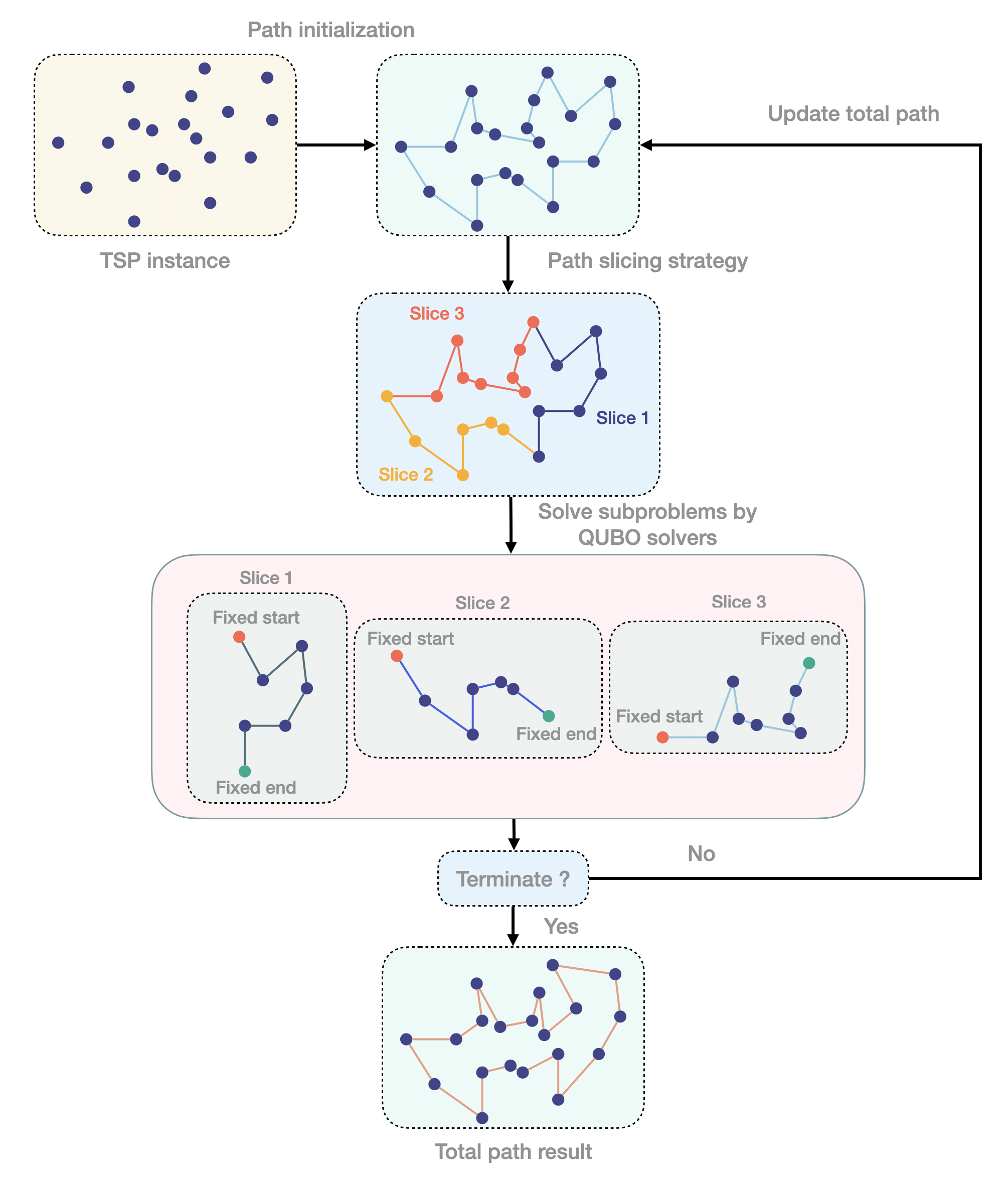}
\caption{Illustration of quantum local search with path slicing strategies. 
}
\label{fig:qls_flow}
\end{figure}

As illustrated in Fig.~\ref{fig:qls_flow}, we begin with a path initialization method to obtain the initial path for a given TSP instance. Following this, we employ the path-slicing strategies discussed later to define subproblems. These subproblems are then addressed using quadratic unconstrained binary optimization (QUBO) solvers, particularly quantum solvers, which may operate in parallel or sequentially. Once we obtain the new sub-path for each slice, we accept the update if it results in a shorter distance and reject it otherwise. This step leads to the formation of a new complete path. 
We iteratively proceed to the iteration step of the quantum local search until the predetermined termination threshold. After iteration, the final path is presented as the result.


\subsection{Random variable picking and path slicing}

\begin{figure}[ht]
\centering
\includegraphics[scale=0.26]{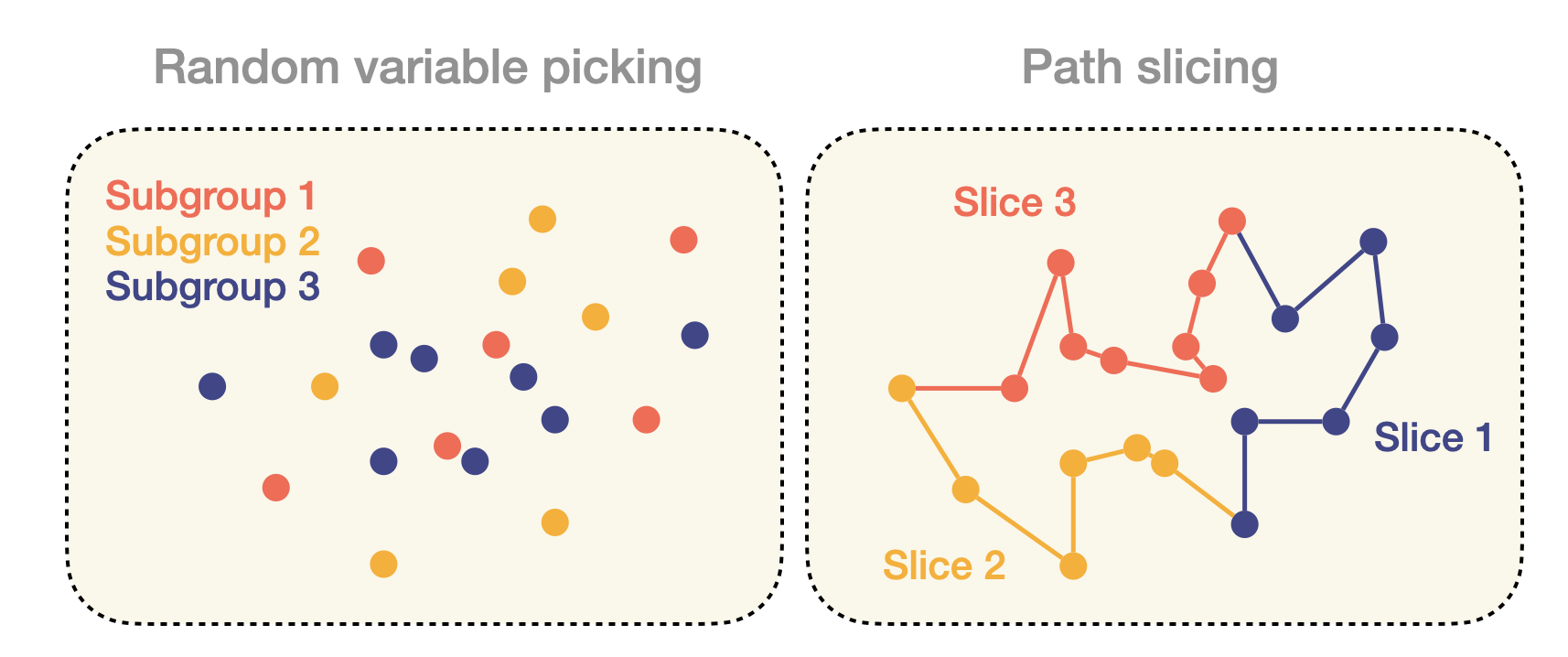}
\caption{Illustration of random variable picking and path-slicing. 
}
\label{fig:rvp_vs_ps}
\end{figure}

We begin by examining the rationale behind employing the path-slicing method for the TSP. The conventional QUBO formulation of the TSP \cite{isingformulation} without the closed path constraint is expressed as follows:
\begin{equation}
    \begin{split}
        H_{ce} =& A\sum_{i=1}^n \left(1 - \sum_{j = 1}^n x_{i,j}\right)^2 +  A\sum_{j=1}^n \left(1 - \sum_{i = 1}^n x_{i,j}\right)^2 \\
    &+\sum_{u,i} D_{ui} \sum_{j=1}^{n-1} x_{u,j} x_{i,j+1},
    \end{split}
    \label{ceTSP}
\end{equation}
where $n$ represents the number of cities, and the binary variable $x_{i,j} \in \{ 0, 1\}$ denotes the presence (or absence) of the salesman at location $i$ during step $j$. $D_{ui}$ indicates the distance between locations $u$ and $i$, while $A$ serves as the penalty for constraint violations. The first and second terms ensure that the salesman visits exactly one location at each step and that each location is visited only once in the entire route. The third term, the objective function, aims to minimize the total distance traveled. The indices $i = 1,..,n$ and $j = 1, ..., n$ result in a total of $n^2$ variables of $x$. It is important to note that in this formulation, a valid path representation using $x$ would appear as a sequence of $n$ one-hot vectors.

Due to the number of variables required for the TSP scaling as \(n^2\), the number of required qubits increases rapidly, making it challenging for the fixed-size hardware of quantum computers to directly encode all variables onto qubits.
In general, for combinatorial optimization problems where the problem size exceeds the solver's capabilities, it is common to employ variable-wise local search techniques such as random variable picking to define subproblems and solve them iteratively. If the solver is a quantum solver, this approach is referred to as quantum local search \cite{qls1, qls2, qls3, qls4, qls5}. It is logical to assert that different types of combinatorial optimization problems may benefit from distinct local search strategies that perform more effectively. 
This study proposes a quantum local search method incorporating specific path-slicing strategies for TSP.

As depicted in Fig.~\ref{fig:rvp_vs_ps}, we graphically compare different methods of forming subproblems within a complete TSP. We have solved eq.~\eqref{ceTSP} to find the optimal path of the subproblem. With random variable picking, once we have formed three subgroups, these can each generate their own sub-paths, maintaining a certain degree of flexibility to reassemble these sub-paths into a complete path \cite{qls6}. In contrast, with the path-slicing method \cite{ps1}, after forming an initial closed path that visits all locations, we can divide the path into segments to optimize the distance within each segment. This approach not only simplifies the process of reconstructing the full path from the sub-paths but also reduces the number of variables required for each subproblem from \(m^2\) to \((m-2)^2\), where \(m\) represents the number of the cities in each segment since the start point and the end point are fixed. 

\subsection{Convex hull insertion heuristic initialization}

Continuing from the previous section, to effectively employ the path-slicing method, we first need an initial path of the complete TSP. In this hybrid quantum-classical heuristic approach, we do not solely rely on quantum solvers to address the TSP; instead, we utilize a classical method for path initialization. This differs from the conventional quantum annealing method, where the initial state of the spin configurations is prepared in the \(|+\rangle^{\otimes N}\) state (the ground state of \(\sum_{i=1}^N \sigma_x^{(i)}\)), where $N$ is the number of qubits, and $\sigma_x^{(i)}$ is the Pauli $x$ operator for site $i$. Our initial path is generated using a classical technique, specifically the convex hull insertion heuristic \cite{chinit_1}, which creates paths with relatively short distances compared to random paths. However, these paths still exhibit a gap from the minimal distance path. Despite this, the method offers a favorable computational complexity of \(O(n^2)\), where \(n\) is the number of locations \cite{chinit_2, chinit_3}.

\subsection{K-means strategy and anti-k-means strategy }
\begin{figure}[ht]
\centering
\includegraphics[scale=0.25]{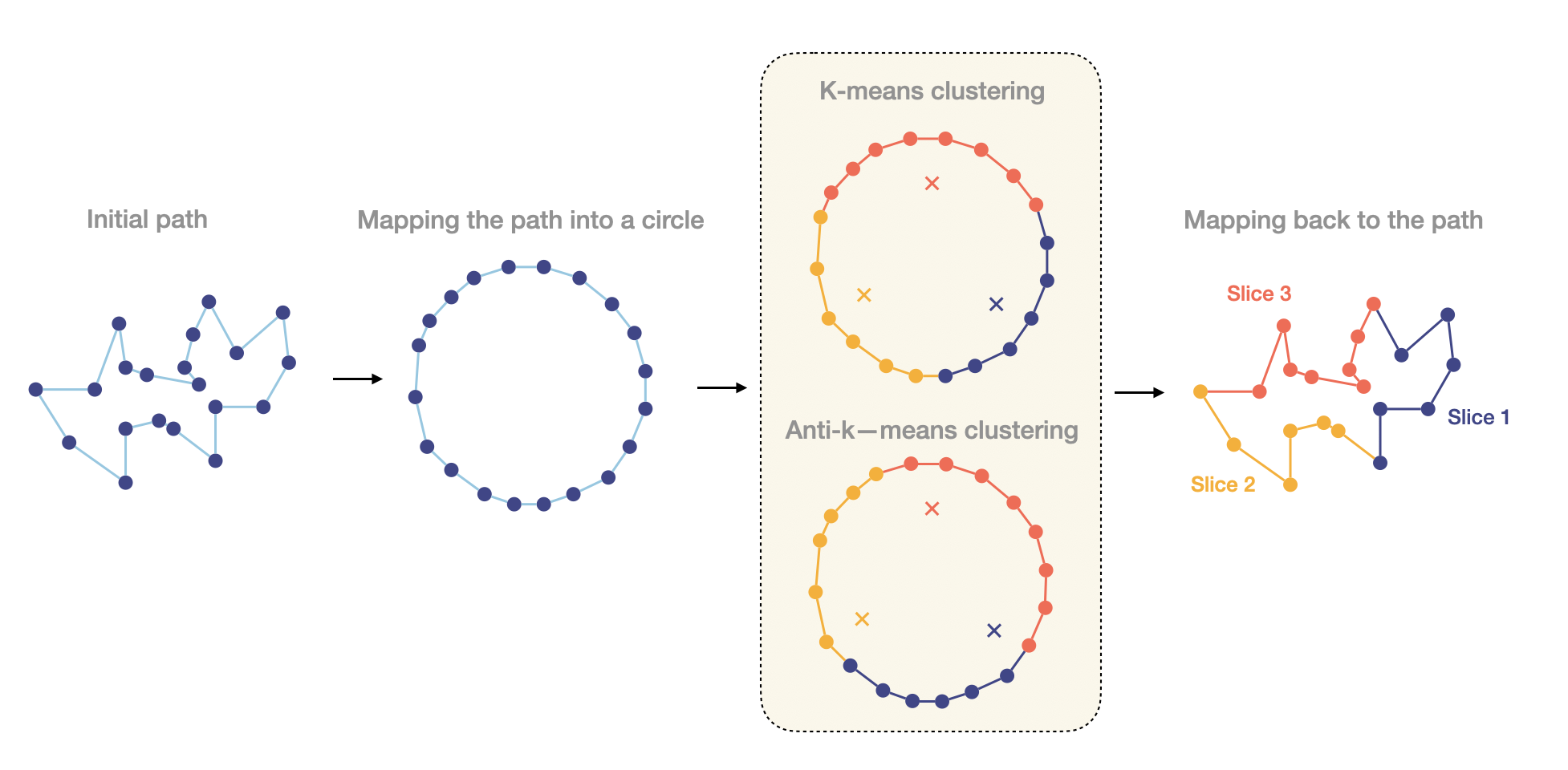}
\caption{Illustration of slicing strategies: k-means and anti-k-means. 
}
\label{fig:ps_strategy}
\end{figure}

After establishing an initial path, we explore the strategies employed in this study. 
In the path-slicing method, the current path should be divided into several clusters.
We have employed k-means clustering to divide the current path.
K-means clustering is a popular unsupervised learning algorithm used to partition data into $k$ distinct clusters, minimizing the variance within each cluster~\cite{kmeans1}. The intuition behind k-means is to find the centroids of the clusters and assign each city to the nearest centroid.
The previous work also employs k-means clustering to TSP~\cite{kmeans2, kmeans3}.
A potential issue with directly applying the k-means method to a TSP instance is that, although it is feasible, there is a possibility that it might create groups where optimal solutions are unattainable (Figure 5 in \cite{kmeans2}) and may not align with our predetermined path.

In this study, we transformed the data before applying k-means.
As illustrated in Fig.~\ref{fig:ps_strategy}, we first transform the information about the relative distances between neighboring locations into a circular format, where the arc distances reflect their proximities in the original path. 
For each position in the path with $n$ locations, we have the distance data between the connected points: $[d_0, d_1, \ldots, d_{n-1}]$. Assuming a circle radius of $r = 1$, we calculate the total circumference $C = \sum_{i=0}^{n-1} d_i$. To place these data points on a circle, we calculate the angular displacement for each segment:
\begin{eqnarray}
\Delta \theta_i = 2\pi \left(\frac{d_i}{C}\right) \quad \text{for } i = 0, 1, \ldots, n-1.
\end{eqnarray}
The cumulative angles $\theta_i$ are then calculated as follows:
\begin{eqnarray}
&& \theta_0 = 0, \\
&& \theta_i = \sum_{j=0}^{i-1} \Delta \theta_j \quad \text{for } i = 1, 2, \ldots, n .
\end{eqnarray}
Next, we convert each angle to the corresponding $(x,y)$ coordinates on the circle:
\begin{eqnarray}
&& x_{\text{circle}, i} = r \cos (\theta_i) \quad \text{for } i = 1, \ldots, n, \\
&& y_{\text{circle}, i} = r \sin (\theta_i) \quad \text{for } i = 1, \ldots, n.
\end{eqnarray}
The resulting arrays $x_{\text{circle}}$ and $y_{\text{circle}}$ contain the coordinates of the points mapped onto the circle.

Once the data is configured in a circular format, the k-means method is applied to segment the data into subproblems. Upon mapping the locations back to the original path, we derive \(k\) slices of the TSP path. In contrast to traditional k-means, where each cluster centroid denotes the center of a slice, we also introduce a comparative strategy named anti-k-means. This method utilizes the k-means centroids as breakpoints for the slices, assuming that in some cases, this will be a more effective way to partition the path, as depicted in Fig.~\ref{fig:ps_strategy}. We will later see that these strategies yield better results under different settings.

\subsection{Hybrid strategy and hybrid-anti strategy}
Although the strategies described above seem promising, they can encounter challenges during quantum local search, particularly with both potentially becoming trapped in local minima within the cluster settings. Here, the clustering—achieved through k-means or anti-k-means—might reach a satisfactory score but still falls short of adequately solving the TSP(s). To address this, we propose a mechanism that allows the cluster configuration to escape these local minima. Consequently, during the quantum local search iterations, we introduce random slicing clusters intermittently with k-means or anti-k-means iterations. In these random slicing clusters, a random displacement between the slice indices is applied, effectively generating random slices. The randomly produced slices provided a ``jump'' between k-means or anti-k-means iterations. For instance, the slicing method for quantum local search iterations could alternate between random and k-means slices—termed a hybrid strategy—as in ``random-(k-means)-random-(k-means)-...", and similarly, the alternative using anti-k-means would follow a sequence of ``random-(anti-k-means)-random-(anti-k-means)-...", which we refer to as the hybrid-anti strategy.


\section{Result and Discussion}
\label{sec:result}
\begin{figure*}[ht]
\centering
\includegraphics[scale=0.26]{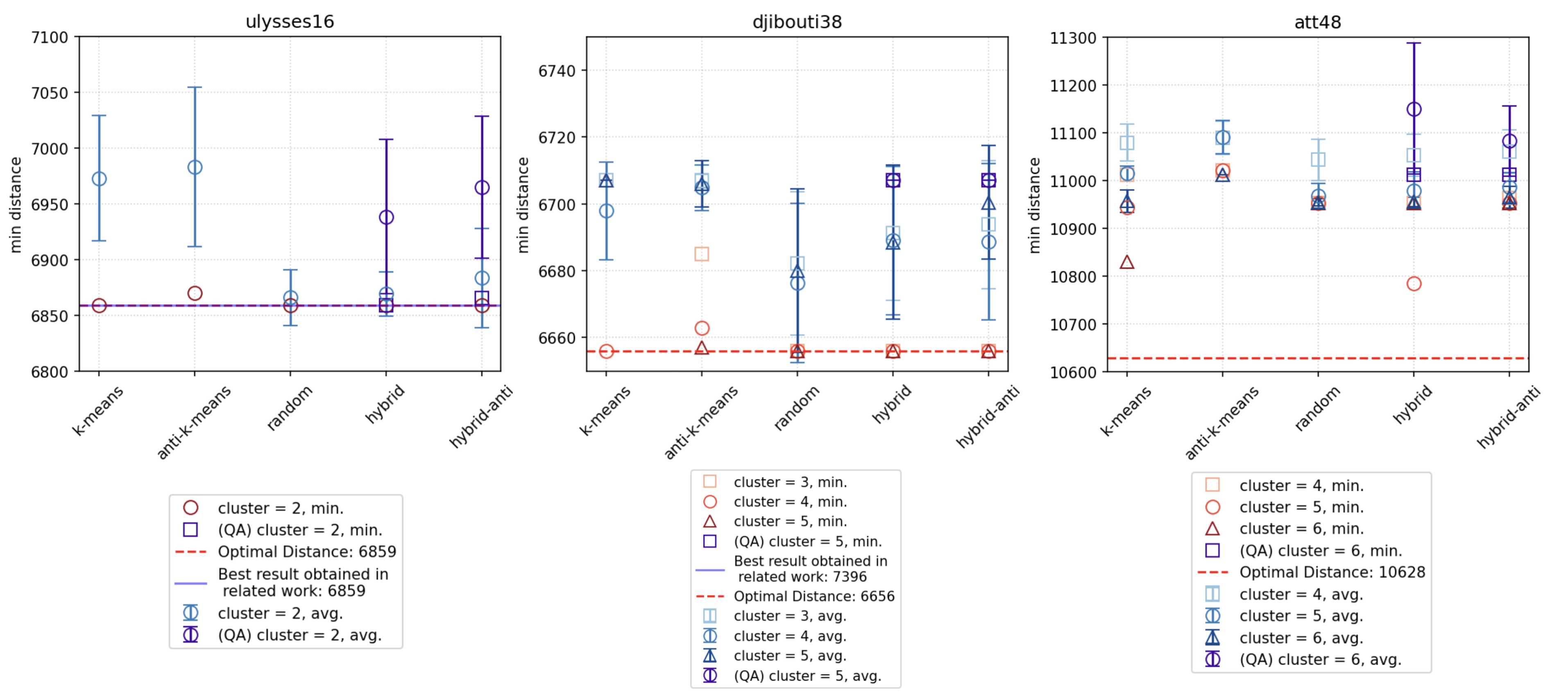}
\caption{Comparison of slicing strategies for the TSP across different instances: ulysses16, djibouti38, and att48. 
}
\label{fig:res_1}
\end{figure*}
 To align our study with the state-of-the-art, we have selected TSP instances from TSPlib\cite{tsplib1}, similar to those used in \cite{cvrp1}. In addition to the 16-location instance, ulysses16, and the 38-location instance, djibouti38, we also explore a larger instance not examined in \cite{cvrp1}, namely the 48-location instance att48, which represents the capitals of 48 US states. With examples of using both simulated annealing and quantum annealing (D-Wave quantum annealer) for subproblems, we demonstrated that our method is a general framework capable of integrating any solver, both quantum and classical, provided they can tackle QUBO problems.
 

As shown in Fig.~\ref{fig:res_1}, we present a comparative analysis of various slicing strategies applied to the TSP across three instances. The strategies examined include k-means, anti-k-means, random, hybrid, and hybrid-anti approaches. The figures illustrate the minimum and average distances achieved with different cluster configurations (represented by different symbols) for each method based on 100 experiments for simulated annealing and 10 experiments for quantum annealing, where the error bars are calculated by standard deviation. Each experiment incorporates 100 iterations of quantum local search. The red dashed line marks the optimal distance for each instance, serving as a benchmark for assessing the effectiveness of each strategy, while the blue lines indicate the best results obtained in previous studies \cite{cvrp1}.

For the ulysses16 instance, both simulated and quantum annealing with a hybrid strategy achieved the optimal distance of 6859.
In the simulated annealing case, the random slicing method appears to be the most effective strategy with less variance.
For the djibouti38 instance, the random strategy consistently outperformed other strategies on average in the simulated annealing case. In this instance, we also explored different numbers of clusters, finding no clear trend indicating the optimal cluster size. Remarkably, we managed to achieve the optimal distance of 6656 for djibouti38, whereas the related work reported the best distance of 7396. 
In the quantum annealing case, we could not obtain the optimal distance by both hybrid and hybrid-anti strategies. However, our best distance obtained is 6707. This is still better than the related work. It should be noted that this value is mainly contributed by the convex hull initialization. The initial path is so good that quantum annealing could not find a better path during the optimization process. 

In the att48 instance, both quantum and simulated annealing achieved almost the same quality in the best solution.
This result indicates that quantum annealing can improve the initial path in this instance.
The hybrid strategy proved to be effective, although it did not reach the optimal distance. The best result was quite close, at 10784 compared to the optimal 10628. Interestingly, in this case, a larger number of clusters—which correspond to smaller subproblems—led to better performance, except for the anti-k-means strategy.

In Table~\ref{tab:res_2}, we present the required subproblem sizes associated with varying numbers of clusters for different TSP instances. The subproblem size is determined by the results of k-means/anti-k-means clustering used in quantum local search, with a larger number of clusters resulting in smaller subproblems. 
In these cases, the reduction in variables is significant. Notably, referring to the maximum subproblem sizes—which could correlate to the number of qubits required by quantum hardware—we reduced qubit usage from 256 to 64 for ulysses16, from 1444 to 196 for djibouti38, and from 2304 to 144 for att48. While methods for reducing qubit requirements are well established, the high quality of the solutions we obtained is particularly noteworthy. We achieved optimal distances in instances where previous work fell short, and we could tackle larger problem instances with quite impressive solution quality. 
This result is achieved by combining the classical initialization strategy, which can already find a good path in some instances, with simulated or quantum annealing, which is then used to attempt to improve the solution quality from this ‘already good’ path.


\begin{table*}[]
\centering
\caption{Comparison of required subproblem sizes for the TSP across different instances: ulysses16, djibouti38, and att48}
\begin{tabular}{|l|c|ccc|ccc|}
\hline
Problem Instance                & ulysses16 & \multicolumn{3}{c|}{djibouti38}                                 & \multicolumn{3}{c|}{att48}                                       \\ \hline
Full problem required variables & 256       & \multicolumn{3}{c|}{1444}                                       & \multicolumn{3}{c|}{2304}                                        \\ \hline
n\_cluster                      & 2         & \multicolumn{1}{c|}{3}     & \multicolumn{1}{c|}{4}     & 5     & \multicolumn{1}{c|}{4}      & \multicolumn{1}{c|}{5}     & 6     \\ 
QLS subproblem size mean     & 37.3      & \multicolumn{1}{c|}{119.8} & \multicolumn{1}{c|}{66.25} & 40.24 & \multicolumn{1}{c|}{115.25} & \multicolumn{1}{c|}{67.88} & 44.63 \\
QLS subproblem size min      & 16        & \multicolumn{1}{c|}{49}    & \multicolumn{1}{c|}{9}     & 4     & \multicolumn{1}{c|}{25}     & \multicolumn{1}{c|}{4}     & 1     \\
QLS subproblem size max      & 64        & \multicolumn{1}{c|}{225}   & \multicolumn{1}{c|}{225}   & 196   & \multicolumn{1}{c|}{361}    & \multicolumn{1}{c|}{169}   & 144   \\ \hline
\end{tabular}

\label{tab:res_2}
\end{table*}




\section{Conclusion}
\label{sec:conclusion}

This work has provided an exploration of the path-slicing strategies applied to the TSP using quantum local search methods. Our approach leverages a combination of quantum and classical techniques to efficiently solve subproblems derived from the TSP, addressing the limitations posed by the current era of NISQ hardware, particularly the restrictions on qubit numbers.

Our investigation into various path-slicing methods, including k-means, anti-k-means, hybrid, and hybrid-anti strategies, has demonstrated their efficacy in optimizing TSP solutions. The results from applying these strategies across different TSP instances, such as ulysses16, djibouti38, and att48, indicate significant improvements in finding near-optimal paths with fewer computational resources compared to existing methods.

A key finding of this research is the effectiveness of random slicing in combination with structured clustering methods like k-means and anti-k-means, which helps prevent the search from becoming trapped in local minima, a common issue in quantum local search applications. The hybrid and hybrid-anti strategies, in particular, have shown promising results by introducing variability and flexibility into the search process, thereby enhancing the overall search dynamics.

Moreover, applying our quantum local search method has reduced the quantum resource requirements. By efficiently managing qubit utilization, our method has extended the practical applicability of quantum computing solutions to larger TSP instances that were previously challenging due to hardware limitations.

The proposed path-slicing strategies for quantum local search contribute to the field of combinatorial optimization by providing a scalable, efficient, and effective approach to solving TSPs. These strategies not only address the inherent limitations of quantum hardware but also set a foundation for future research in the optimization of other complex problems using hybrid quantum-classical approaches. The results encourage ongoing and future studies to refine these techniques further and explore their application in broader contexts, pushing the boundaries of what is computationally feasible with emerging quantum technologies.



\bibliographystyle{IEEEtran}
\bibliography{references}

\end{document}